\documentclass[pre,epsfig,address,twocolumn,showpacs]{revtex4}%
\usepackage[english]{babel}
\usepackage{amsmath}
\usepackage{color}
\usepackage{epsfig}
\usepackage{graphicx}
\usepackage{bm}
\usepackage{times,amsmath,amssymb}
\usepackage{subfigure}
\usepackage{amsfonts}
\usepackage{amssymb}%
\begin{document}
\title{Full decoherence induced by local fields in open spin chains with
strong boundary couplings
}
\author{Vladislav Popkov$^{1,6}$,  Mario Salerno$^2$, and Roberto Livi$^{1,3,4,5}$}
\affiliation{$^{1}$ Dipartimento di Fisica e Astronomia, Universit\`a di Firenze, via G.
Sansone 1, 50019 Sesto Fiorentino, Italy \\
$^{2}$ Dipartimento di Fisica "E. R. Caianiello", CNISM and INFN Gruppo
Collegato di Salerno, Universit\`{a} di Salerno, via Giovanni Paolo II
Stecca 8-9, I-84084, Fisciano (SA), Italy.\\
$^3$ Max Planck Institute for the
Physics of Complex Systems,
N\"othnitzer Stra\ss e 38,
01187 Dresden
Germany\\
$^4$ INFN, Sezione di Firenze, and CSDC Universit\`a di Firenze, via G.Sansone 1, 50019 Sesto Fiorentino, Italy\\
$^5$ ISC-CNR, via Madonna del Piano 10, 50019 Sesto Fiorentino, Italy\\
$^6$ Institut f\"{u}r Theoretische Physik, Universit\"{a}t zu K\"{o}ln, Z\"ulpicher Str. 77,
50937 Cologne, Germany
}
\begin{abstract}
We investigate an open $XYZ$ spin $1/2$ chain driven out of equilibrium by
boundary reservoirs targeting different spin orientations, aligned along
the principal  axes of anisotropy.
We show that by tuning  local magnetic fields, applied to spins at  sites near the
boundaries, one can change any nonequilibrium steady state
to a fully uncorrelated Gibbsian state at infinite temperature.
This phenomenon occurs for strong boundary coupling and on a
critical manifold in the space of the fields amplitudes. The structure of this manifold depends on the
anisotropy degree of the model and on the parity of the chain size.
\end{abstract}

\maketitle

\section{Introduction}
Manipulating a quantum system in non--equilibrium conditions
appears nowadays one of the most promising perspectives for proceeding our
exploration of the intrinsic richness of quantum physics and for obtaining an insight
on its potential applications \cite{BlatterPhysReports08,BlattNature2013,ZollerNature2008}.
In particular, much attention has been devoted to
the study of the nonequilibrium steady state (NESS) in quantum spin chains,
coupled to an environment, or a measuring apparatus.
This is described, under Markovianity assumptions \cite{Petruccione,PlenioJumps,ClarkPriorMPA2010}, in the framework
of a Lindblad Master equation (LME) for a reduced density matrix, where a unitary evolution,
described via Hamiltonian dynamics, is competing with a Lindblad dissipative action.
Under these conditions,  quantum spin chains subject to a gradient
evolve towards a NESS, where spin and
energy currents set in. In quasi one-dimensional systems, such currents
exhibit quite exceptional properties like scalings, ballisticity and integrability
\cite{reviewBrenig07,KlumperLectNotes2004,ProsenExact2011,ZnidaricHeisIsotr2011,MPA2013,CaiBarthel13,HeatTransfer}.
 Many of these unexpected features
stem from the fact that the NESS, corresponding to a fixed point of the LME dissipative
dynamics with a gradient applied at the chain boundaries,  are not  standard Gibbs-states.
Moreover, further peculiar regimes
appear when the time lapse between two successive
interactions of the quantum chain with the Lindblad reservoir becomes
infinitely small, while the interaction amplitude is properly rescaled.
In the framework
of projective measurements, this kind of experimental protocol corresponds to
the so-called Zeno effect, that determines how frequent projective measurements
on a quantum system have to be performed in order to {\sl freeze} it in a given state
\cite{Zeno,ZenoAntizeno}.

In this paper we shall rather focus  on a  Zeno regime for non-projective measurements,
that has been found to describe new counterintuitive scenarios for NESS. In particular,
in  \cite{VNE-XXZ}  it was shown that in a boundary driven
XXZ  spin chain,  for  suitable values of the spin anisotropy
the NESS is a pure state.
We want to point out the importance of this result in the perspective of engineering
dark states, that have the advantage  to be more stable
against decoherence, than isolated quantum systems and, therefore,  better candidates
for technological applications \cite{ZollerPRA08,ZollerNature2008}. Here we investigate how this {\sl non projective}
Zeno regime can be manipulated
by the action of a  strictly local magnetic field, whose strength is of the order of the
exchange interaction energy of the XYZ Heisenberg spin chain model. The main result of our investigations
is that, by such a local effect, one can kill any coherence of the NESS and turn it into
a mixed state at infinite temperature. More generally, the von Neumann entropy of the NESS
can be changed from its minimum value to its maximum one just by tuning
the local magnetic field, provided  the coupling with the baths is sufficiently strong.


The paper is organized as follows. In Section \ref{sec::The model} we describe the
general properties of the {\sl non projective} Zeno setup and the way the spin XYZ
chain is coupled to the Lindblad reservoirs. The effect of complete decoherence induced by the addition of a fine-tuned
local magnetic field acting on the spins close to the boundaries are discussed in
Section \ref{sec::Manipulations of NESS by non uniform external fields}.
A short  account of the symmetries characterizing
the NESS in the special case of a XXZ spin chain is reported in Section \ref{sec::Symmetries of NESS}.
In Section \ref{sec::Noncommutativity} we investigate the non--commutativity of the different
limits to be performed in the model and the presence of corresponding hierarchical singularities.
We conclude with a discussion on the perspectives of our investigations  (see Section \ref{sec::Conclusion}).

Appendices \ref{app::Inverse of the Lindblad dissipators and secular conditions},\ref{app:Analytic treatment of $N=3$ case},
\ref{app::Proof of (ii)} and \ref{app::Proof of hierarchical singularity in the NESS for N=4} contain
some relevant technical aspects.

\section{The model}

\label{sec::The model}

We study an open chain of $N$ quantum spins, represented by the
Hamiltonian operator  $H$ , in contact with boundary
reservoirs. The time evolution of the reduced density matrix $\rho$ is
described by a quantum Master equation in the Lindblad form \cite{Petruccione,PlenioJumps,ClarkPriorMPA2010} (we set $\hbar=1$)

\begin{equation}
\frac{\partial\rho}{\partial t}=-i\left[  H,\rho\right]  +\Gamma(%
\mathcal{L}%
_{L}[\rho]+%
\mathcal{L}%
_{R}[\rho]), \label{LME}%
\end{equation}
where $\mathcal{L}_{L}[\rho],\mathcal{L}_{R}[\rho]$ are Lindblad dissipators acting on spins at the left and right boundaries of the chain,
respectively.
This is  an usual setup for studying  transport in quantum spatially extended systems,  where the
explicit choice of $\mathcal{L}_{L}$ and $\mathcal{L}_{R}$
is suggested by the kind of application one has in mind. In this way, one describes an effective
coupling of the  chain, or a part of it, with  baths or
environments. Within the quantum protocol of repeated interactions,
Eq.(\ref{LME}) describes an
exact time evolution of the extended quantum system, provided  the coupling with the Lindblad reservoirs
is suitably rescaled \cite{ClarkPriorMPA2010}.


Here we are interested to explore the strong coupling condition, i.e.  $\Gamma\rightarrow\infty$,
that corresponds to the so--called  {\sl Zeno regime}. In this case one can obtain the
stationary solution of Eq.(\ref{LME})  in the form of the perturbative
expansion
\begin{equation}
\rho_{NESS}(\xi,\Gamma)=\sum_{k=0}^{\infty}\left(  \frac{1}{2\Gamma}\right)
^{k}\rho_{k}(\xi), \label{PT_largeCouplings}%
\end{equation}
where $\rho_{NESS}(\xi,\Gamma)$ is the density matrix of the non equilibrium steady--state
and  $\xi$ is a symbol epitomizing the model parameters (e.g. bulk anisotropy, exchange
energy, magnetic field, etc.).

Suppose that the stationary solution $\rho_{NESS}%
(\xi,\Gamma)$ is unique. This fact will be validated further for all our examples.
Moreover, the  first term of
expansion (\ref{PT_largeCouplings}), i.e.  $\rho_{0}=$ $\lim_{\Gamma
\rightarrow\infty}\lim_{t\rightarrow\infty}\rho(\Gamma,\xi,t)$, satisfies  the
stationarity condition $%
\mathcal{L}%
_{LR}[\rho_0]=0$, where  $%
\mathcal{L}%
_{LR}=%
\mathcal{L}%
_{L}+%
\mathcal{L}%
_{R}$ is the sum of the Lindblad actions in (\ref{LME}) . This suggests
that $\rho_{0}$  can be represented in a factorized form
\begin{equation}
\rho_{0}=\rho_{L}\otimes\left(  \left(  \frac{I}{2}\right)  ^{\otimes_{N-2}%
}+M_{0}(\xi)\right)  \otimes\rho_{R}, \label{InitialChoiceRo0}%
\end{equation}
where $\rho_{L}$ and $\rho_{R}$ are the one-site density matrices at the
chain boundaries, satisfying  $\mathcal{L}_{L}[\rho_{L}]=0$ and  $\mathcal{L}_{R}[\rho_{R}]=0$,
 and $M_{0}$ is a matrix to be determined self-consistently.
It is convenient to separate explicitly the identity matrix $\left(  \frac{I}{2}\right)  ^{\otimes_{N-2}}$
from $M_{0}$, in such a way that  $M_{0}$  is a traceless operator, due to the
condition $Tr(\rho_{0})=1$ .

By substituting the perturbative expansion (\ref{PT_largeCouplings}) into Eq.(\ref{LME}) and
by equating terms of the order
$\Gamma^{-k}$,  one can easily obtain the recurrence relation
\begin{equation}
i[H,\rho_{k}]=\frac{1}{2}%
\mathcal{L}%
_{LR}\rho_{k+1}\text{ , \ }k=0,1,2, \cdots \label{Recurrence0}%
\end{equation}
whose general solution has the form
\begin{equation}
\rho_{k+1}=2%
\mathcal{L}%
_{LR}^{-1}(i[H,\rho_{k}])+\rho_{L}\otimes M_{k+1}\otimes\rho_{R}\text{,
\ \ \ }k=0,1,2, \cdots \label{Recurrence}%
\end{equation}
provided  the following secular conditions (for more details see \cite{XYtwist}) are satisfied
\begin{equation}
\lbrack H,\rho_{k}]\cap\ker(
\mathcal{L}_{LR})=\varnothing , \label{SecularConditionsGeneral}
\end{equation}
where $\ker(\mathcal{L}_{LR}) $ denotes the nucleus of the operator $\mathcal{L}_{LR}$.

Notice that, in order to obtain an
explicit solution, one has to  compute the inverse operator $ \mathcal{L} _{LR}^{-1}$,
that appears in Eq.(\ref{Recurrence}).

In summary, Eqs (\ref{InitialChoiceRo0}),
(\ref{Recurrence}) and (\ref{SecularConditionsGeneral}) define a general
perturbative approach,  that applies in the Zeno (i.e., strong coupling) regime.

We   consider the Hamiltonian
\[
H=H_{XYZ}+V_{2}+V_{N-1}
\]
where
\begin{equation}
H_{XYZ}=\sum_{j=1}^{N-1}\left( J_x \sigma_{j}^{x}\sigma_{j+1}^{x}+ J_y\sigma_{j}%
^{y}\sigma_{j+1}^{y}+ \Delta \sigma_{j}^{z}\sigma_{j+1}^{z}\right)  ,
\label{Hamiltonian}%
\end{equation}
is the Hamiltonian of an open $XYZ$ Heisenberg spin chain and $V_{l}$ is
a local inhomogeneity field acting on spin $l$ to be specified later on (see Eqs \ref{Vds2}-\ref{VdsNm1}).
Moreover, we consider Lindblad dissipators, $\mathcal{L}_{L}$ and  $\mathcal{L}_{R}$,
favouring a relaxation of boundary spins at $k=1$
and $k=N$ towards states described by one-site density matrices $\rho_{L}$ and
$\rho_{R}$, i.e.  $\mathcal{L}_{L}[\rho_{L}]=0$ and  $\mathcal{L}_{R}[\rho_{R}]=0$.
In particular,  we choose boundary reservoirs that tend to align the spins
at the left  and right  edges along the  directions $\vec{l}_{L}$ and $\vec{l}
_{R}$, respectively. These directions are identified by the longitudinal and azimuthal  coordinates as follows:
\[
\vec{l}_{L,R}=(\sin\theta_{L,R}\cos\varphi_{L,R},\sin\theta_{L,R}\sin
\varphi_{L,R},\cos\theta_{L,R}).
\]

Such a setting is achieved by choosing the Lindblad action in the form
$\mathcal{L}[\rho]=
\mathcal{L}_{L}[\rho]+\mathcal{L}_{R}[\rho]$, where%
\begin{equation}
\mathcal{L}_{A}[\rho]=-\frac{1}{2}\left\{  \rho,\mathcal{S}_{A}^{\dag}\mathcal{S}_{A}\right\}
+\mathcal{S}_{A} \rho\,
\mathcal{S}_{A}^{\dag}, \,\,\,\,\,\,\, A = \, L\, , \, R
\label{LindbladActionArbitrary}%
\end{equation}
and
\begin{align}
\mathcal{S}_{L}  &  =[(\cos\theta_{L}\cos\varphi_{L})\sigma_{1}^{x}+(\cos\theta_{L}%
\sin\varphi_{L})\sigma_{1}^{y}-(\sin\theta_{L})\sigma_{1}^{z}+\nonumber\\
&  i\sigma_{1}^{x}(-\sin\varphi_{L})+i\sigma_{1}^{y}(\cos\varphi_{L})]/2,
\label{L_Left}%
\end{align}%
\begin{align}
\mathcal{S}_{R}  &  =[(\cos\theta_{R}\cos\varphi_{R})\sigma_{N}^{x}+(\cos\theta_{R}%
\sin\varphi_{R})\sigma_{N}^{y}-(\sin\theta_{R})\sigma_{N}^{z}+\nonumber\\
&  i\sigma_{N}^{x}(-\sin\varphi_{R})+i\sigma_{N}^{y}(\cos\varphi_{R})]/2.
\label{L_Right}%
\end{align}
In the absence of the unitary term in (\ref{LME}), the boundary spins relax
with a characteristic time $\Gamma^{-1}$ to specific states described via the
\textit{one-site} density matrices
\begin{align}
\rho_{L}  &  =\frac{1}{2}\left(  I+\vec{l}_{L}\vec{\sigma}_{1}\right)
\label{RoL}\\
\rho_{R}  &  =\frac{1}{2}\left(  I+\vec{l}_{R}\vec{\sigma}_{N}\right)  .
\label{RoR}%
\end{align}
The condition $%
\mathcal{L}%
_{A}[\rho_{A}] = 0
$ follows from  definition (\ref{LindbladActionArbitrary}),
while the relations $\mathcal{S}_{A}\mathcal{S}_{A}^{\dagger}=\rho_{R}$ and $(\mathcal{S}_{A})^{2}%
=(\mathcal{S}_{A}^{\dagger})^{2}=0$ can be easily checked.

In analogy with  \cite{ProsenUniqueness}, it can be easily  shown  that, for the chosen boundary dissipation
setup described by Eqs (\ref{LindbladActionArbitrary}), (\ref{L_Left}) and (\ref{L_Right}), the NESS is unique.
By applying the perturbative approach in the Zeno regime, one
finds  that the unknown
matrices $M_{k}(\Delta)$ are fully determined by secular conditions
(\ref{SecularConditionsGeneral}). As shown in Appendix \ref{app::Inverse of the Lindblad dissipators and secular conditions}, for the specific choice  (\ref{LindbladActionArbitrary}) of the
Lindblad operators,  they are equivalent to the requirement of a null partial trace
\begin{equation}
Tr_{1,N}([H,\rho_{k}])=0\text{, \ }k=0,1,2, \cdots  \label{SecularConditions} \,\,\,\,\,\, .
\end{equation}

We want to point out that the computation of the full set of matrices $\{M_{k}(\Delta)\}$ for any $\Delta\neq0$  is quite
a nontrivial task. However, in the Zeno limit, $\Gamma \rightarrow \infty$, we are just interested in
computing the zero--th
and the first order contributions $M_{0},M_{1}$, which  can be completely
determined by solving the set of secular equations (\ref{SecularConditions}) for $k=0,1,2$.

\section{Manipulations of NESS by non uniform external fields}
\label{sec::Manipulations of NESS by non uniform external fields}

The  properties of the model introduced in the previous section have been widely investigated for $V_l = 0$ and $\varphi=\pi/2$  in \cite{XYtwist}. Here we are interested in studying how the properties of the NESS can be modified when $V_l$ is an additional local field, that corrupts the homogeneity of the XYZ spin chain.

Notice first that  a local field applied to the boundary spins at positions $k=1$ and $k=N$
does not affect the strong coupling limit $\rho_{0}=$ $\lim_{\Gamma\rightarrow\infty}\rho_{NESS}(\Gamma
)$. On the other hand, applying a local field to the spins at positions
$ k=2$ and $k=N-1$ can modify $\rho_{0}$ in a nontrivial way.
The Hamiltonian  reads
\begin{equation}
\label{HT2}
H=H_{XYZ} + V_{2} + V_{N-1}
\end{equation}
where
\begin{align}
V_{2}  &= \vec{h}\vec{\sigma}_{j} = h_{x}\sigma_{2}^{x}+h_{y}\sigma_{2}^{y}+h_{z}\sigma_{2}^{z}
\label{Vds2} \\
V_{N-1}  &= \vec{g}\vec{\sigma}_{N-1} = g_{x}\sigma_{N-1}^{x}+g_{N-1}\sigma_{j}^{y}+g_{z}\sigma_{N-1}^{z}
\label{VdsNm1}
\end{align}

Carrying out the procedure outlined in the previous section, we can find
the form of the density matrix of the NESS\ in the Zeno regime,  $\rho_{0}$. This is a function
of the angles $\theta_{L},\varphi_{L},\theta_{R},\varphi_{R}$, of the anisotropy
parameter  $\Delta$ and of the local fields $\vec{h},\vec{g}$.
One can argue that, in general, the NESS should be an  entangled state, depending in a
nontrivial way on  the local fields. Due to the boundary drive, the NESS typically exhibits nonzero
currents (magnetization current, heat current, etc.), irrespectively of the presence of the local fields. However, in the  Zeno limit, there are critical values of the local fields for which  a complete decoherence of the NESS occurs.

More precisely,  we formulate our results under the following boundary condition assumptions:
\begin{itemize}
\item targeted boundary polarizations are neither collinear nor anti-collinear (${\vec l}_L \neq \pm {\vec l}_R$);
\item at least one of the polarizations (e.g. the left targeted polarization) is directed along one
of the anisotropy axis $X,Y,$ or $Z$;
\item the corresponding local fields ($\vec{h}$  at site 2  and $\vec{g}$  at site $N-1$) are collinear to the respective targeted
boundary polarizations  $\vec{h}=h {\vec l}_L$, $\vec{g}=g {\vec l}_R$.
\end{itemize}
 Then, there exists  a zero-dimensional or a one dimensional critical manifold in  the $h,g$--plane
 $(h_{cr},g_{cr})$, such that, in the Zeno limit , the NESS on this manifold becomes
 \begin{equation}
\left.  \rho_{NESS}(\Delta)\right\vert _{(h_{cr},g_{cr})}
=\rho_{L}\otimes\left(  \left(  \frac{I}{2}\right)  ^{\otimes_{N-2}}\right)
\otimes\rho_{R},\label{GibbsInfiniteTemperature}%
\end{equation}
Notice that this a peculiar state:  apart  the frozen boundary spins, all the internal spins are at
infinite temperature.
 Indeed, tracing out the boundary spins, one obtains the Gibbs state at infinite temperature
 \begin{equation}
Tr_{1,N}\left(  \rho_{L}\otimes\left(  \left(  \frac{I}{2}\right)  ^{\otimes_{N-2}}\right)
\otimes\rho_{R} \right)=  \left(  \frac{I}{2}\right)^{\otimes_{N-2}}.
\label{Tr1,N}
\end{equation}
Also notice that on the critical manifold the Von-Neumann entropy of the NESS, $S_{VNE}=-Tr(\rho_{NESS}\log_{2}\rho_{NESS})$, in the Zeno limit attains its maximum
value given by
\[
\lim_{\Gamma\rightarrow\infty}\max(-Tr(\rho_{NESS}\log_{2}\rho_{NESS}))=N-2,
\]
since $\rho_{L},\rho_{R}$ are pure states.
In the following, we also refer to  state (\ref{GibbsInfiniteTemperature}) as the {\sl state of  maximal decoherence}.

We have performed explicit calculations (see below)
that confirm the above  statement for different spin chains up to   $N=8$. The particular form of the NESS assumed in these cases, however,  strongly suggests that the above results maybe of  general validity and  the critical manifold $(h_{cr},g_{cr})$  independent on  $N$.

The critical manifold has been fully identified for the following cases.
\vskip .2 cm
- \textit{XYZ chain}:  $J_x \neq J_y \neq \Delta$.
If the left, ${\vec l}_L$, and the right, ${\vec l}_R$,  polarizations point in directions of different principal axes
\begin{equation}
{\vec l}_L = e_\alpha \,\,\,\, , \,\,\,\, {\vec l}_{R} = e_\beta  \,\,\,\,\, \alpha\not=\beta  \,\,\,\,\, \alpha, \beta = X, Y, Z
\end{equation}
where $e_X=(1,0,0),e_Y=(0,1,0), e_Z=(0,0,1)$,
then for chains with an even number, $N$, of spins, the manifold $(h_{cr},g_{cr})$  consists of three critical points:
$P_{\alpha} = (-2 J_\alpha,0)$, $P_{\beta} = (0, -2 J_\beta)$ and $P_{\alpha,\beta} = (-J_\alpha, - J_\beta)$. For odd $N$, the critical point  $P_{\alpha,\beta}$ is missing and the critical manifold reduces only to the points $P_{\alpha}, P_{\beta}$, above. If only one of the two
boundary driving points in the direction of a principal axis, the critical manifold  reduces to a single point, either $P_{\alpha}$ or $P_{\beta}$, for both even and odd $N$.

\vskip .2 cm
- \textit{XXZ chain}: $J_x = J_y=J \neq \Delta$. If both
$\vec{l}_L$ and  $\vec{l}_R$ lay onto the $XY$--plane,
we can parametrize  the
targeted boundary polarizations via a twisting angle in the $XY$--plane
$\varphi$ as $\theta_{1}=\theta_{2}=\frac{\pi}{2},\varphi_{1}=\varphi,\varphi_{2}=0$, corresponding to
$\vec{l}_{L}=(\cos\phi,\sin\phi,0)$ and $\vec{l}_{R}=(1,0,0)$.
\begin{figure}[ptb]
\centerline{
\includegraphics[width=0.4\textwidth]{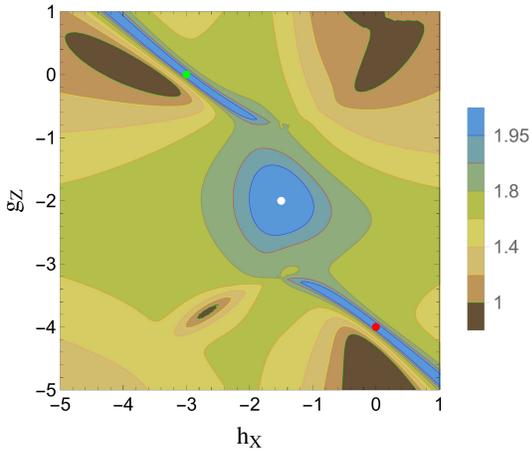}}
\caption{(Color online)
Contour plot of the Von-Neumann entropy $S_{VNE}$ in the Zeno limit, as a function of the local fields for an open  XYZ chain of $N=4$ spins with exchange parameters $J_x=1.5  J_y=0.8, \Delta=2$. Green, white and green dots denote the critical points $P_{X}=(-2 J_X, 0)$, $P_{XZ}=(-J_X, -\Delta)$,   $P_{Z}=(0, -2 \Delta)$, where the VNE reaches its maximum value $S_{VNE}=2$ and  the NESS becomes a completely mixed state, respectively. Other parameters are fixed as $\vec l_L=e_X$, $\vec l_R=e_Z$.
Green, yellow, pink, orange, brown, red and blue contour lines refer to $S_{VNE}$ values: $1, 1.2, 1.4, 1.6, 1.8, 1.9, 1.95$, respectively. Notice the presence of the narrow corridors around $P_{X}$ and
$P_{Z}$ in which the deviation, $S_{VNE}-2$, of the VNE from its maximum value becomes very small.
}
\label{fig1}
\end{figure}
The critical fields are aligned parallel to the targeted boundary magnetization, i.e.
$\vec{h}_{cr}=(h_{cr}\cos\phi,h_{cr}\sin\phi,0)\,\, , \,\, \vec{g}_{cr}=(g_{cr},0,0)$,
and we find the one--dimensional critical  manifold
\begin{equation}
h_{cr}+g_{cr}   =-2 J, \ h_{cr}\neq -J
\label{CriticalParametersArbXYtwist}
\end{equation}
Notice that this expression is independent of system size $N$,
of the anisotropy $\Delta$ and of the twisting angle $\varphi$.
If one of the two targeted polarizations points out of the $XY$--plane,  the critical  manifold
becomes zero-dimensional  and consists of one, two or three critical points (depending on the polarization direction and on $N$ being even or odd) as discussed for the full anisotropic case.

\vskip .2 cm
- \textit{XXX chain}:  $J_x = J_y = \Delta \equiv J $.  The  critical manifold for
arbitrary non-collinear boundary drivings is one-dimensional and it is given
by Eq.(\ref{CriticalParametersArbXYtwist}).

\vskip .2 cm

\begin{figure}[ptb]
\centerline{
\includegraphics[width=0.4\textwidth]{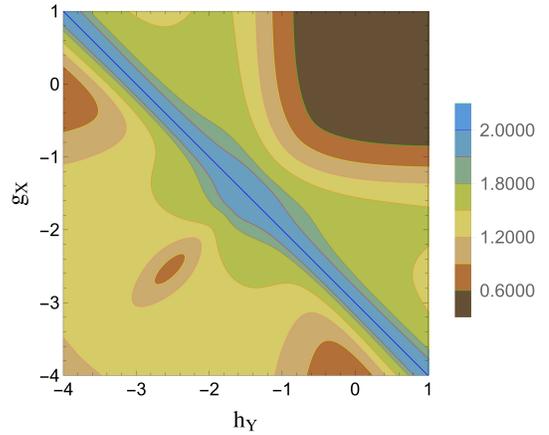}
}
\caption{(Color online) Contour plot of $S_{VNE}$ in the Zeno limit, as a function of the local fields for the XXZ chain with $N=4$ spins. Parameters are  $J_x=J_y\equiv J=1.5, \, \Delta=1, \vec l_L=-e_Y$, $\vec l_R=e_X$. The green, yellow, pink, orange, brown, red, blue contour lines refer to $S_{VNE}$ values: $0.6, 0.9, 1.2, 1.5, 1.8, 1.9, 2$, respectively. The blue contour is in full overlap with the critical line $h_Y+g_X=-2 J=-3$.}
\label{fig2}
\end{figure}

The above statements are illustrated  in Figs. \ref{fig1} and \ref{fig2} for the case of a chain of N=4 spins. In particular, in Fig. \ref{fig1} we show a contour plot of the VNE surface as a function of the applied fields for the  XYZ case with  left and right boundary polarizations fixed along the X and Z directions, respectively. The three critical points $P_{X}, P_{Z}, P_{X,Z}$ mentioned above correspond to the  green, red, and white dots shown in the top panel of the Figure. Notice the presence of narrow corridors (blue shaded) around the $P_X$ and $P_Z$ critical points, inside which  the VNE keeps very close to the maximal value $S_{VNE}=2$ but never reach it, except at the critical points. This is quite different from the partially anisotropic  XXZ case shown in Fig. \ref{fig2}, where the existence of the critical line (blue line) is  quite evident.

\begin{figure}[ptb]
\centerline{
\includegraphics[width=0.4\textwidth]{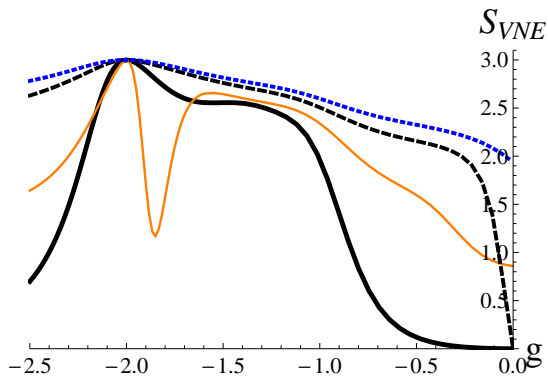}
}
\caption{(Color online) Von-Neumann entropy of the NESS $S_{VNE}=-Tr(\rho_{NESS}\log_{2}\rho_{NESS})$
 in the Zeno limit, as function of
local field $g$, for different values of spin exchange anisotropy.
Thick,thin,dashed  and dotted curves correspond to $\Delta= 0.9239,0.6,0.3827,0.3$, respectively.
For $g=-2$ the NESS is a completely mixed state for which VNE reaches its upper
limit. Parameters: $h=0,\ N=5,\theta_{L}=\theta_{R}=\frac{\pi}{2},\varphi_{L}=-\pi/2,\varphi_{R}=0$.
}
\label{FigN5VNE}
\end{figure}

Similar results are found also  for longer chains. In particular, in Fig.\ref{FigN5VNE} we show a cut of the VNE surface for a partially anisotropic XXZ chain of $N=5$ spins.  For the sake of simplicity we have set $J_x = J_y=1$ and considered the cut at $h=0$ so that the VNE of the NESS, in the Zeno limit, becomes  a function of  $g$ only. We see that for  $g=g_{cr}=-2$, the VNE reaches the maximum value $N-2$ indicating that the corresponding NESS  has the form (\ref{GibbsInfiniteTemperature}).

\begin{figure}[ptb]
\centerline{
\includegraphics[width=0.4\textwidth]{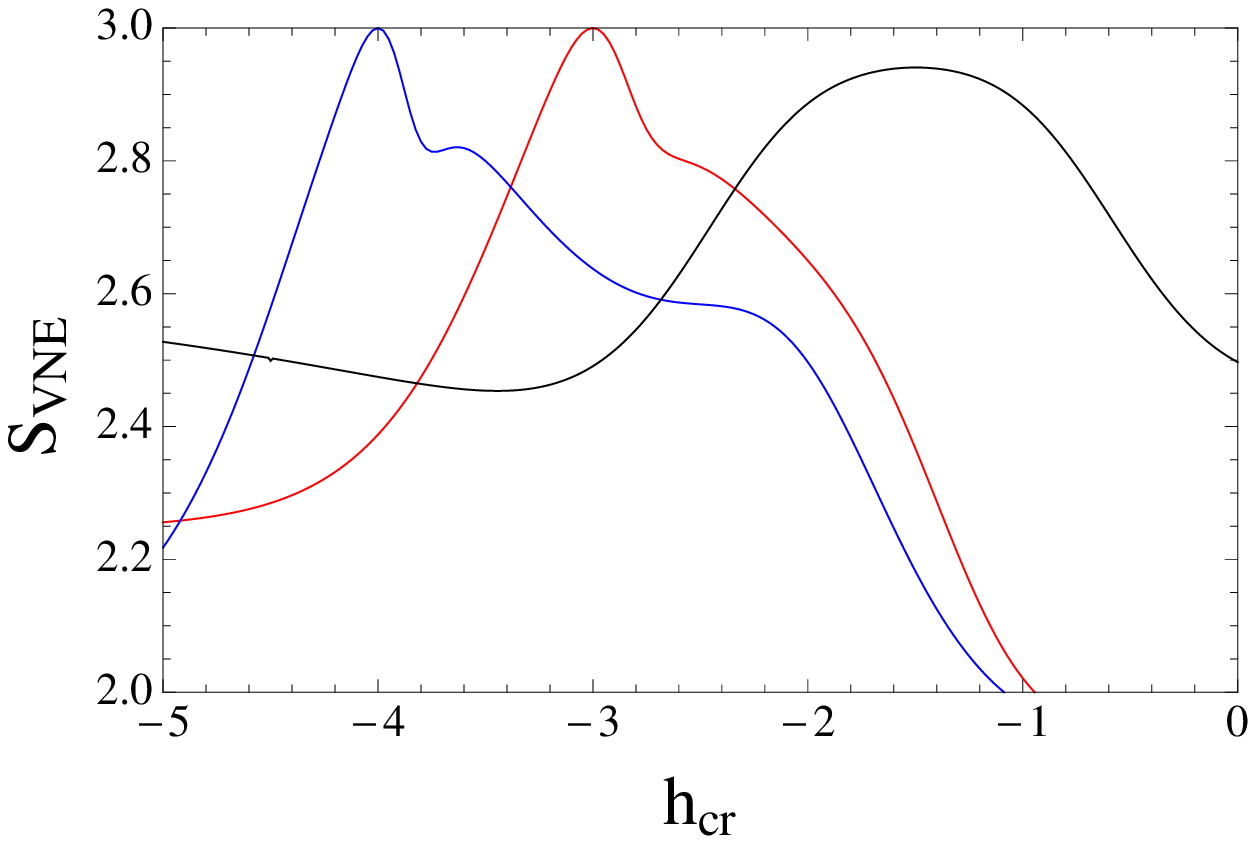}}
\centerline{
\includegraphics[width=0.4\textwidth]{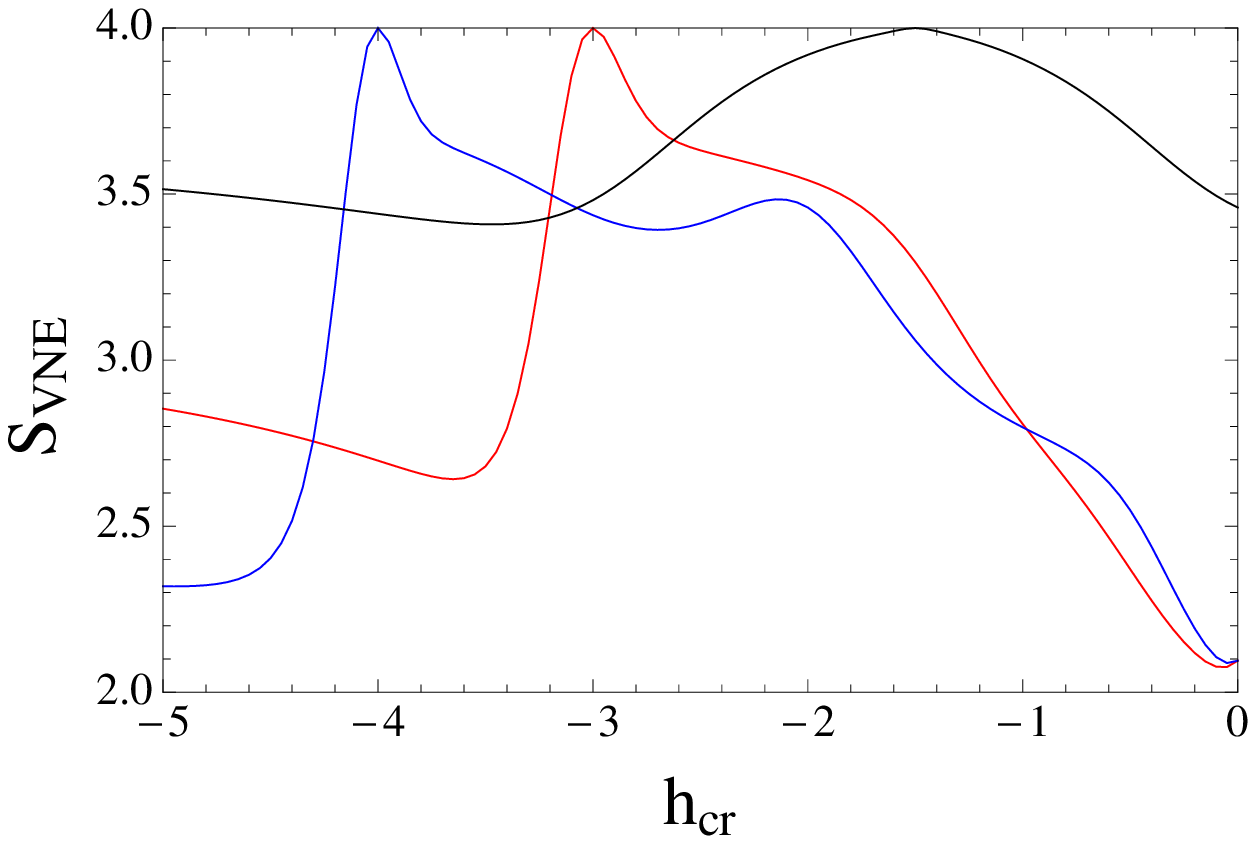}
}
\caption{(Color online) Cuts of the Von-Neumann entropy surface of the NESS
 in the Zeno limit, as function of
critical  field  for the XYZ chains with  $N=5$ (top panel) and $N=6$ (bottom panel) spins.
The red, blue  and black lines refer to cuts made at $g_Z=0$, $h_X=0$ and $g_Z=-2$, respectively.
Other parameters are fixed  as in Fig. \ref{fig1}. Notice that for $N$ odd the VNE reaches its maximum  value $N-2$ only at points $P_X=(-2 J_x, 0)$ and  $P_Z=(0,-2 \Delta)$ while for $N$ even the maximum is reached also at the point  $P_{XZ}=(-J_x, -\Delta)$.
}
\label{fig3}
\end{figure}

As to the dependence of the critical manifold on parity of $N$, we find
that while for odd sizes $N=3,5,7$ and $XYZ$ Hamiltonian ( see  Fig. \ref{fig3}, top panel for an illustration) there are only two critical points (the critical point $P_{\alpha,\beta}$ is missing), for  even sizes $N=4,6,8$ cases there are  three critical points. These observations strongly suggest a qualitative difference between even and odd $N$ in the
model, which is manifested in other NESS properties as well, see e.g. (\ref{SymmetryEven}),(\ref{SymmetryOdd}).
%

It is worth to note here that for $h=g=-J$, i.e. the case excluded in (\ref{CriticalParametersArbXYtwist}), the NESS
behaves non-analytically in the Zeno limit  $\Gamma \rightarrow \infty$.
As we are going to discuss in Sec.\ref{sec::Noncommutativity}, this non-analyticity
is a consequence of  the non-commutativity of the limits
$\Gamma \rightarrow \infty$ and $h=g  \rightarrow -J$.

\begin{figure}[htbp]
\begin{center}
 {\includegraphics[width=0.4\textwidth]{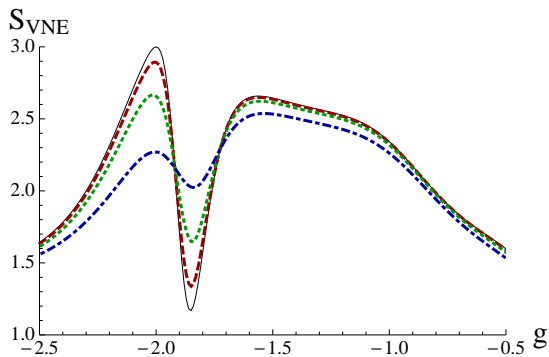}}
 \caption{(Color online)
 Von-Neumann entropy of the NESS  as function of the
local field $g$ and for different values of the coupling $\Gamma$. Other parameters are fixed as:
 $N=5$, $\Delta=0.6, J_x =J_y=1$,  $h=0,\theta_{L}=\theta_{R}=\frac{\pi}{2},\varphi_{L}=-\pi/2,\varphi_{R}=0$.
The thin (black),  red (dashed), dotted (green), dot-dashed (blue) curves refer to values $\Gamma= 10^3, \Gamma= 10^2, \Gamma=50, \Gamma=25$, respectively.
}
\label{msfig4}
\end{center}
\end{figure}

Conversely, for any finite boundary coupling $\Gamma$, i.e. far from  the Zeno limit, the  NESS is analytic for
arbitrary amplitudes of the local fields (the first order correction to the NESS for large  $\Gamma$ is proportional to $\Gamma^{-1}$ as shown in the App.\ref{app::Proof of (ii)}).
This is also seen from  Fig.\ref{msfig4} where
the VNE of the NESS is reported as a function of the
local field $g$  for different values of the boundary coupling $\Gamma$ and same parameters as in Fig.\ref{FigN5VNE} (see curve $\Delta=0.6$). Notice that the thin black line obtained for $\Gamma=10^3$, is already in  full overlap with  the  Zeno limiting curve depicted in Fig.\ref{FigN5VNE} for $\Delta=0.6$.  Also note the  persistence of the peak at $g=-2$ even for relatively small values of $\Gamma$ away from the Zeno limit.

Similar behaviors are observed  for different choices of boundary polarizations and of local fields (not shown for brevity), thus opening the  possibility to detect the signature of the above  phenomena in real experiments.
In this respect we remark that the near--boundary magnetic field $h$ and the anisotropy $\Delta$
as suitable parameters for controlling the dissipative state of the system in a NESS.
Thus, if  $g=0$, the NESS can be made a pure state by  tuning the anisotropy $\Delta$ to a specific value
$\Delta^{\ast}(\varphi,N)$. For instance, we find that for $g=0$ and
    $\Delta_{\pm}^{\ast}(\pi/2,5)=\sqrt{\frac{1}{2} \pm \frac{1}{2 \sqrt{2}}}$ the NESS is a pure state
\cite{PureStateN5delta},
 while  for $g_{cr} =-2$, the NESS in the bulk
becomes an infinite temperature state (\ref{GibbsInfiniteTemperature}), i.e. a maximally mixed state.
Thus, by suitably tuning the
anisotropy and the local magnetic field one can pass from minimally mixed (pure) NESS state to a maximally mixed one.


It should be emphasized at this point that general thermodynamic equilibrium quantities, e.g. the temperature,
are not well-defined for a generic NESS. In fact, pure states allowed by Liouvillian dynamics are not ground
states of the Hamiltonian, but are characterized by a property of being common eigenvectors of a
modified Hamiltonian and of all Lindblad operators \cite{Yamamoto05,ZollerPRA08}. Likewise, an absence
of currents in the NESS does not necessarily imply a thermalization of the system: in fact also for fully
matching boundary conditions the NESS is not a Gibbs state at some temperature,  so that correlation
functions remain far from those of an equilibrium system. From this point of view, the decoherence effect described in
present paper can be viewed as a  reaction of a nonequilibrium system on a local perturbation (the local magnetic field): as is well-known,
a local perturbation in nonequilibrium can lead to global changes of a steady state.

On the other hand, a fully mixed state as such has appeared already in the context of driven spin chains: if both Lindblad boundary reservoirs target trivial states with zero polarization ($\rho_R=\rho_L=I/2$), the NESS is maximally
mixed $\rho_{NESS}= \left(  \frac{I}{2}\right)^{\otimes_{N}}$, which is a trivial solution
of the steady Lindblad equation for any value of  boundary coupling. The respective NESS is often being referred to as a state with infinite temperature \cite{Znidaric2011}.
Note that our case is drastically different from the latter: the maximally mixed state (\ref{GibbsInfiniteTemperature}) appears only in the bulk, after tracing the boundary spins, in a system with generically strong boundary gradients, and under  strong boundary coupling.

A few more remarks are in order:
(i) the amplitudes of the critical local fields scale with the amplitude of
the Hamiltonian exchange interaction, i.e.
$h_{cr}\rightarrow \gamma h_{cr}$ if $H_{XXZ}\rightarrow \gamma H_{XXZ}$
(this is a consequence of  the linearity of the recurrence relations (\ref{Recurrence}) and (\ref{SecularConditions}) );
(ii) the NESS may take the form (\ref{GibbsInfiniteTemperature}) only in the Zeno limit
 $\Gamma\rightarrow\infty$; in fact, the first order correction to the
NESS is proportional to $\Gamma^{-1}$ and does not vanish (see Appendix
\ref{app::Proof of (ii)});  the fully decoherent  state  (\ref{GibbsInfiniteTemperature}) is intrinsic
to nonequilibrium conditions and, strikingly enough, it persists even
for nearly matching or fully matching boundary driving, as we are going to discuss in Sec.\ref{sec::Equilibrium and near equilibrium boundary driving}.

We want to conclude this Section by pointing out that a fully analytic treatment of the problem
for arbitrary large values of $N$
should encounter serious technical difficulties. The main one concerns the solution of
the consistency relations determined by the secular conditions (\ref{SecularConditionsGeneral}) for the perturbative
expansion (\ref{PT_largeCouplings}), with the zero-order term given by (\ref{GibbsInfiniteTemperature}).
Moreover, finding the first order correction to NESS,  proportional to $\Gamma^{-1}$, amounts to  solve a
system of equations, whose number grows exponentially with $N$.
With Matematica we were able to solve that system of equations analytically for $N \leq5$ and
numerically up to  $N\leq 7$.

\section{Matching and quasi--matching boundary drivings}
\label{sec::Equilibrium and near equilibrium boundary driving}

In the previous Section we have discussed the case where the complete alignment
of the boundary Lindblad baths was excluded. In this Section we want to analyze the
specific case where they  are aligned (or quasi--aligned) in the same direction on the $XY$--plane.

A complete alignment, i.e. $\vec{l}_L=\vec{l}_R$, corresponds to a perfect matching
 between the left and right boundary Lindblad baths, that yields a  total absence of boundary gradients,
so that any current of the NESS vanishes. Also in this case the Gibbs state at infinite temperature can be
achieved by suitably  tuning the values of the near--boundary fields, but  for even-sized chains, only.

\begin{figure}[ptb]
\centerline{
\includegraphics[width=0.4\textwidth]{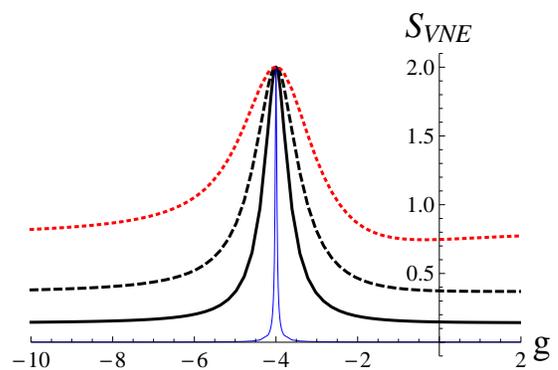}
}
\caption{(Color online) Von-Neumann entropy of the NESS $S_{VNE}=-Tr(\rho_{NESS}\log_{2}\rho_{NESS})$
 in the Zeno limit, as function of
local field $g$, for  $XYZ$ model with matching boundary driving  $\vec{l}_L=\vec{l}_R=(0,0,1)$,
for different values of spin exchange anisotropy difference $J_y-J_x$.
Thin, thick, dashed  and dotted curves correspond to $J_y-J_x= 0.02,0.3,0.6,1.2$.
Parameters:  $J_x=1.5,\Delta=2$, $N=4$.
}
\label{Fig_JxJySingularity4}
\end{figure}

Let us first illustrate this finding for the {\sl XYZ} case.
With no loss of  generality,  we can  set $\vec{l}_L=\vec{l}_R=e_Z=(0,0,1)$.
The behavior of the driven chain with local fields  depends drastically on whether the size of the chain $N$
is an even or an odd number: in the former case we find the critical one dimensional manifold, defined by
 \begin{equation}
 h_{cr}+g_{cr}   =-2 \Delta, \ h_{cr}\neq -\Delta \,\,\ ;
\label{Zcollinear crit point}
\end{equation}
in the latter case $N=3,5,..$, we do not find any critical point. This result has been
found explicitly for  $3 \leq N \leq 6$, but, since it depends on the effect of local perturbations,
it seems reasonable to conjecture that it should hold for larger finite values of $N$.
This result holds as long as the Heisenberg exchange interaction in the plane perpendicular to
the targeted direction
(the $XY$--plane in this example) is anisotropic, i.e.  $J_x \neq J_y$. Conversely, for $J_x = J_y$, the infinite temperature state (\ref{GibbsInfiniteTemperature}) cannot be reached for
any value of the local fields $h$ and $g$. There is a delicate point to be taken into account when
we fix $h = h_{cr}$ and we perform the limit  $J_y \rightarrow J_x$, i.e. we reestablish the model
isotropy: for complete alignment,  $\vec{l}_L=\vec{l}_R=e_Z$, the NESS is singular. The way this
singularity sets in is shown in  Fig. \ref{Fig_JxJySingularity4}.
In the limit when the anisotropy in the direction transversal
to the targeted direction becomes infinitesimally small $|J_y-J_x|\rightarrow 0$
the NESS is a pure state with minimal possible $S_{VNE}\rightarrow 0$ for any amplitude of the local field values, except at a critical point
where  $S_{VNE}$ is maximal.

The  noncommutativity of similar limits and the
dependence of the NESS properties on the parity of system size $N$ in Lindblad--driven Heisenberg chains,
have been observed in previous  studies \cite{SelectionRules2013},\cite{WeakCoupling2013}. Also in these cases,
the origin of noncommutativity is a consequence of  global symmetries of the NESS, that, for our model,
are discussed in Section \ref{sec::Symmetries of NESS}.

In the {\sl isotropic case}, as long as the local fields are parallel to the targeted spin polarization, the NESS does
not depend on them: it is a trivial factorized homogeneous state with a maximal polarization matching the boundaries,
i.e.
 $\rho_{NESS}=(\rho_L)^{\otimes_N}$. This can be easily verified by a straightforward calculation.

Another kind of NESS singularity can be found in the {\sl partially anisotropic} case, with quasi--matching boundary
driving in the $XY$ isotropy plane.  As a mismatch parameter we inroduce
the angular difference between the targeted polarizations at the left and the right boundaries, $ \varphi= \varphi_L- \varphi_R$. For $ \varphi=0$ and in the absence of local fields, we have found that the spin polarization at
each site of the chain is parallel to the targeted polarization; on the other hand, even in the Zeno limit, it does not
saturate to the value imposed at the boundaries $j = 1\, ,\, N$. In general,  this is not an equilibrium Gibbs state,
even in the Zeno limit and for any finite boundary  coupling $\Gamma$.
However,  if the near--boundary fields are switched on and tuned to their critical values, the coherence of this state
is destroyed and the NESS becomes an infinite temperature Gibbs state.
On the other hand,  we have found that there is a relevant difference between  quasi--matching and
mismatching conditions for even and odd values of $N$ (notice that  the isotropic and the free fermion cases,
$\Delta = 1 $,  $\Delta = 0 $,  are special and should be considered separately).
Our results can be summarized as follows:

\begin{figure}[htbp]
\begin{center}
 \subfigure[\label{fig:a}]%
 {\includegraphics[width=0.4\textwidth]{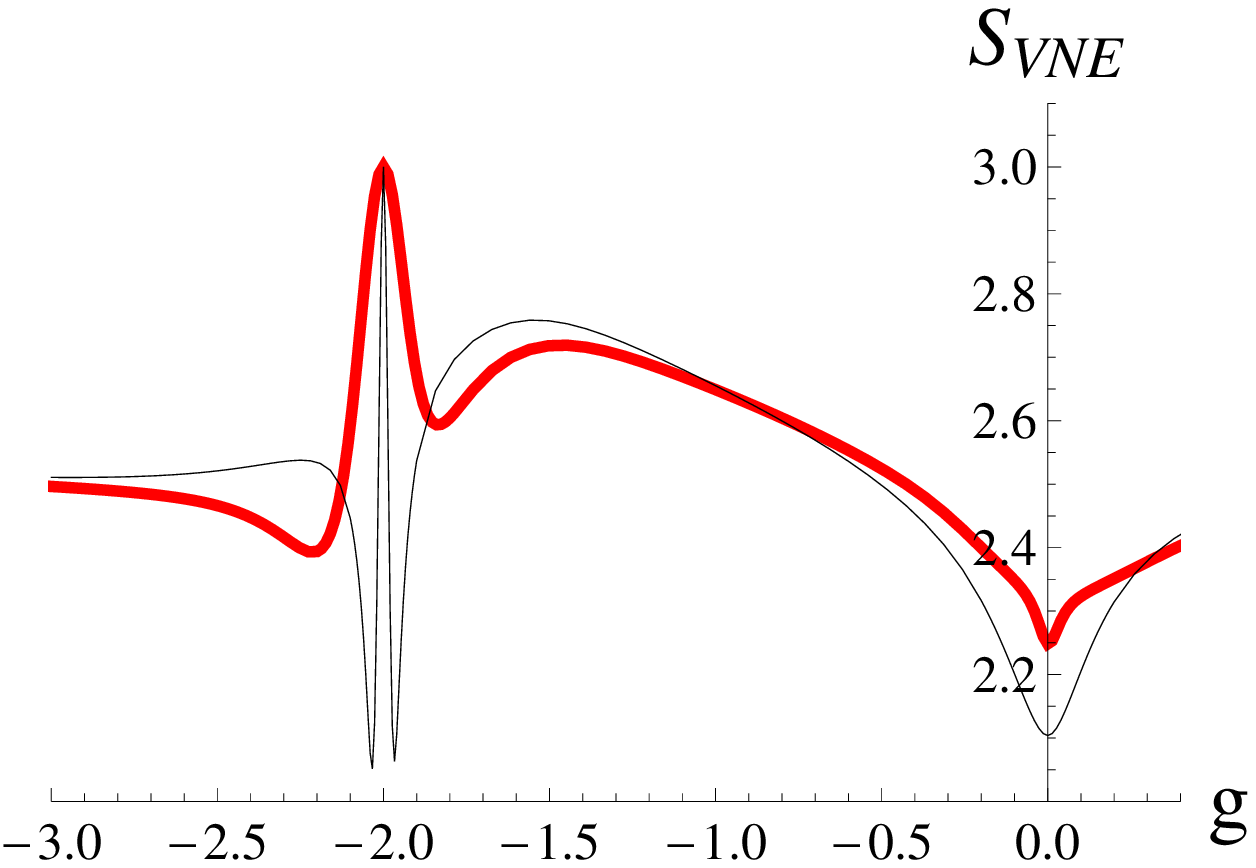}}  \qquad
\subfigure[\label{fig:c}]%
{\includegraphics[width=0.4\textwidth]{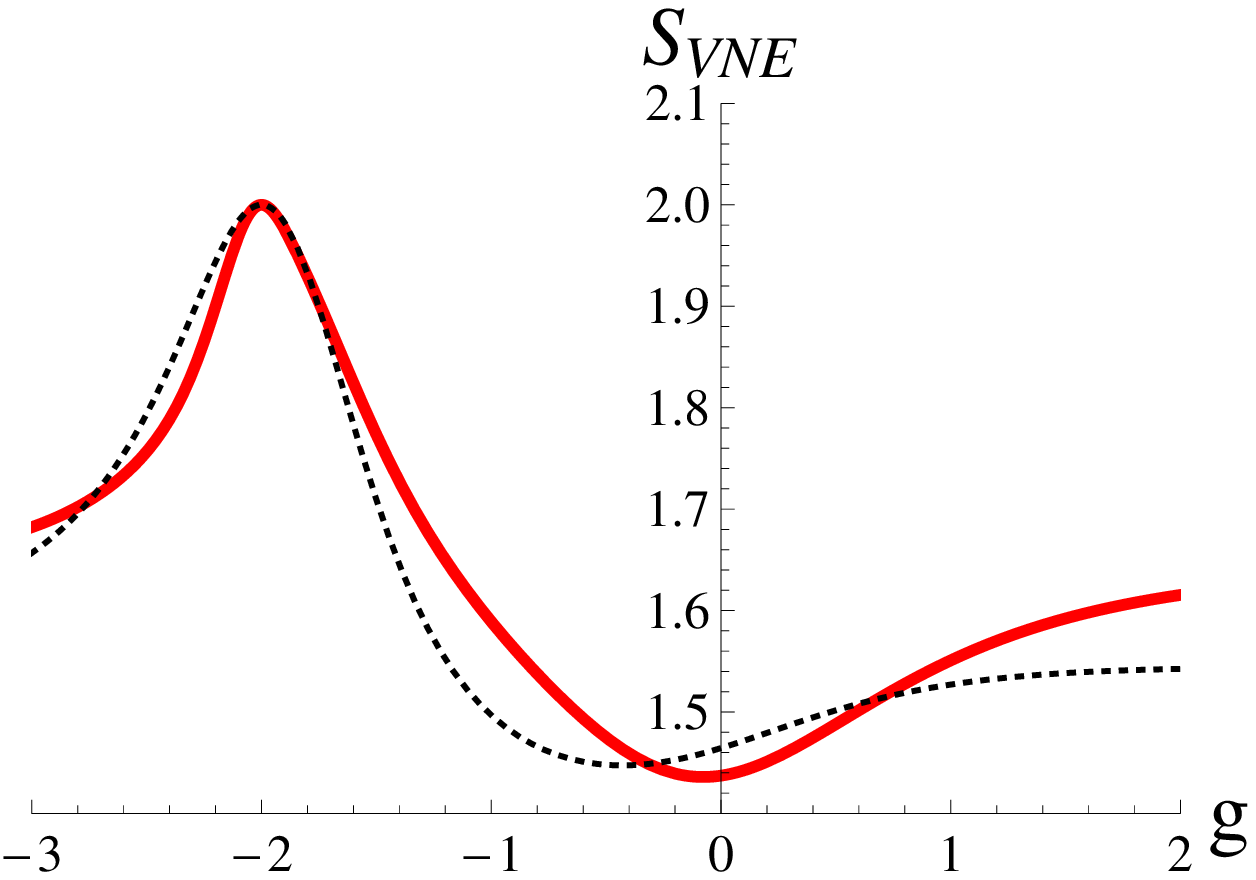}}
\caption{(Color online)
Von Neumann entropy of an internal block, (sites $1,..,N-1$), for $N=5$ (Panel a), and $N=4$ (Panel b),
 versus the local field $g$, for different $\varphi$. Parameters: $\Delta=0.3$.
  Panel (a): Thick and dotted
 curve correspond to $\varphi=\pi/7 $ and  $\varphi=\pi/30$ respectively.
  Panel (b): Thick and dotted
 curve correspond to $\varphi=\pi/7 $ and  $\varphi=0$ respectively.
}
\label{figVNEoddANDeven}
\end{center}
\end{figure}

\vskip .2 cm
 - {\sl $N$ odd.} We can fix the boundary  mismatch by choosing $\varphi_{L}= \varphi, \varphi_{R}=0$, the
 left local field $h=0$, and study the NESS  as a function of the right local field $g$.
 At  $g = g_{cr}=-2$, the NESS becomes trivial (maximally mixed);
 however, as shown in panel (a) of  Fig.\ref{figVNEoddANDeven},  for small mismatch  we find a
 singular behavior of the NESS close to $g = g_{cr}$. Analytic calculations (not reported here) show that
 for $ \varphi=0$ there is a singularity at $g = g_{cr}$, as a result of the  non-commutativity of the limits $ \varphi \rightarrow 0$
 and  $g  \rightarrow g_{cr}$.

\vskip .2 cm
 - {\sl $N$ even.} Unlike the previous case, the NESS is analytic  for
 small  and zero mismatch (see panel (b) of  Fig.\ref{figVNEoddANDeven}).
 For $g = g_{cr}$  the NESS becomes trivial (maximally mixed), also for $ \varphi = 0$.

Finally, let us comment about two special cases,
for  "equilibrium" boundary driving conditions, i.e.  $\varphi_{L}=\varphi_{R}$.
For $\Delta = 0 $ (free fermion case), the NESS is a fully mixed state (apart from the
boundaries) for all values of $g$.
For $\Delta = 1 $ (isotropic Heisenberg Hamiltonian), the NESS is a trivial factorized state, fully polarized
along the axis of the boundary driving, for any  value of $g$.
Both statements can be straightforwardly verified.

NESS singularities, onset  of which can be recognized in Fig.~\ref{Fig_JxJySingularity4} and Fig.~\ref{figVNEoddANDeven}a, appear because of non-commutativity
of limits.
Noncommutativity of various limits, implying singularities and nonergodicity,
which are due to global  symmetries is a well-established phenomenon
and occurs already in Kubo linear response theory describing fluctuations of a
thermalized background. In nonequilibrium open quantum systems, however, the
presence of NESS  symmetries at special value of parameters is manifested much
strongly, due to richer phase space which includes both bulk parameters (such
as anisotropy and external field amplitudes) and boundary parameters (such as
coupling strength). As a result, noncommutativity of the limits and consequent
NESS singularities seems to be a rather common NESS feature.
In the next two sections we reveal some of NESS symmetries and show that the respective
singularities,  connected with them,
can be observed  already in a finite
system consisting of a few qubits.

\section{Symmetries of NESS}
\label{sec::Symmetries of NESS}

Symmetries of the LME are powerful tools that reveal general, system size-independent  properties of the Liouvillean dynamics (\ref{LME}).
In the case of multiple steady states, symmetry based analysis allows one to predict different  basins of attraction of the density matrix  for different initial conditions \cite{MultipleNESSsymmetries14}. For a unique steady state,   symmetry analysis provides a qualitative description of the Liouvillean spectrum \cite{ProsenSymmetries}
or the formulation of selection rules for steady state spin and heat currents \cite{SelectionRules2013}.
It is instructive to list several general NESS symmetries valid for
 our setup. We restrict to $XXZ$ Hamiltonian with $J_x=J_y=1$, and perpendicular targeted polarizations in
the  $XY$--plane, i.e.
 ${\vec l}_L =(0,-1,0)$, ${\vec l}_R =(1,0,0)$.
The
LME has a symmetry, depending on parity of  $N$, which connects the NESS for positive and negative
$\Delta$. Let us denote by $\rho_{NESS}(N,\Delta,h,g,\Gamma)$ the
nonequilibrium steady state solution of the Lindblad master equation (see
(\ref{LME}) and (\ref{LindbladActionArbitrary}) ) for the Hamiltonian (\ref{HamXXZ+LocalFields})
reported in Appendix B.
It is known that this NESS is unique\cite{ProsenUniqueness} for any set of its parameters;
moreover, one can easily check that
\begin{equation}
\rho_{NESS}(N,-\Delta,h,g,\Gamma)=U\rho_{NESS}^{\ast}(N,\Delta,h,g,\Gamma
)U \label{SymmetryEven}%
\end{equation}%
\begin{equation}
\rho_{NESS}(N,-\Delta,h,g,\Gamma)=\Sigma_{y}U\rho_{NESS}^{\ast}(N,\Delta
,h,g,\Gamma)U\Sigma_{y} \label{SymmetryOdd}%
\end{equation}
These relations hold for even and odd values of $N$, respectively; here
$\Sigma_{y}=(\sigma^{y})^{\otimes_{N}}$,
$U={\textstyle\prod\limits_{n\text{ odd}}}\otimes\sigma_{n}^{z}$
and the asterisk on the r.h.s. of both equations
denotes complex conjugation in the
basis where $\sigma^{z}$ is diagonal. Eqs (\ref{SymmetryEven}%
) and (\ref{SymmetryOdd})  hold for any value of the  local fields $h,g$ and for
any coupling $\Gamma$, including the Zeno limit $\Gamma \rightarrow \infty$.
Due to  properties (\ref{SymmetryEven}) and (\ref{SymmetryOdd}), we can restrict
to the case $\Delta\geq0$ further on.
For $g=-h$,  $\rho_{NESS}(N,\Delta,h,g,\Gamma)$
has the automorphic symmetry,
\begin{eqnarray}
&& \rho_{NESS}(N,\Delta,h,-h,\Gamma)= \nonumber \\
&& \Sigma_{x}U_{rot}R\rho_{NESS}(N,\Delta
,h,-h,\Gamma)RU_{rot}^{+}\Sigma_{x}, \label{SymmetryGlobal}%
\end{eqnarray}
where $R(A\otimes B\otimes...\otimes C)=(C\otimes....\otimes B\otimes A)R$ is
a left-right reflection, $U_{rot}=diag(1,i)^{\otimes_{N}}$ is a rotation in
$XY$ plane, $U_{rot}\sigma_{n}^{x}U_{rot}^{+}=$ $\sigma_{n}^{y}$,
$U_{rot}\sigma_{n}^{y}U_{rot}^{+}=-\sigma_{n}^{x}$, and $\Sigma_{x}%
=(\sigma^{x})^{\otimes_{N}}$.

\section{Non-commutativity of the limits $\Gamma\rightarrow\infty$ and
$h\rightarrow h_{crit}$ , $\Delta\rightarrow\Delta_{crit}$. Hierarchical
singularities.}

\label{sec::Noncommutativity}
Here we consider the $XXZ$ Hamiltonian and a perpendicular targeted polarizations in the
 $XY$--plane
 $\vec{l_L}=(0,-1,0)$, $\vec{l_R}=(1,0,0)$;
the  near--boundary fields are taken on the critical manifold, i.e.
 $h+g=-2$.
For $N=3, 5$  and  $\Delta>0$ we have found the noncommutativity conditon
\begin{eqnarray}
& & \lim_{\Gamma\rightarrow\infty}\lim_{h\rightarrow1}\rho_{NESS}(N,h,-h-2,\Delta
,\Gamma) \neq  \nonumber  \\
& & \lim_{h\rightarrow1}\lim_{\Gamma\rightarrow\infty}\rho
_{NESS}(N,h,-h-2,\Delta,\Gamma).
\label{NonCommutativityGeneralDelta}
\end{eqnarray}
Making use of (\ref{GibbsInfiniteTemperature}), the r.h.s. of
(\ref{NonCommutativityGeneralDelta}) can be rewritten
\begin{equation}
\lim_{h\rightarrow1}\lim_{\Gamma\rightarrow\infty}\rho_{NESS}(N,h,-h-2,\Delta
,\Gamma)=\rho_{L}\left(  \frac{1}{2}I\right)  ^{\otimes_{N-2}}\rho_{R}.
\end{equation}
For the simplest nontrivial case $N=3$, the validity of these  noncommutativity
relations is verified by the calculations reported in Appendix B (see (\ref{NonCommutativityN3})).
On top of  (\ref{NonCommutativityGeneralDelta}), we find an
additional singularity at the isotropic point $\Delta=1$ for $N>3$

\begin{eqnarray}
& & \lim_{\Gamma\rightarrow\infty}\lim_{\Delta\rightarrow1}\lim_{h\rightarrow
1}\rho_{NESS}(N,h,-h-2,\Delta,\Gamma) \; \neq \nonumber\\
& & \lim_{\Gamma\rightarrow\infty} \lim_{h\rightarrow1}\lim_{\Delta\rightarrow1}\rho_{NESS}(N,h,-h-2,\Delta
,\Gamma).
\label{NonCommutativityDelta1}
\end{eqnarray}
Due to the symmetry conditions (\ref{SymmetryEven}) and (\ref{SymmetryOdd}), the singularity is
present also for  $\Delta=-1$. Eqs (\ref{NonCommutativityGeneralDelta}%
) and (\ref{NonCommutativityDelta1}) entail the  presence in our model of a
\textit{hierarchical} singularity. Namely, the full parameter space of a model
is a four dimensional one and consists of the parameters $\{\Delta,\Gamma
^{-1},h,g\}$.
As a consequence of  (\ref{NonCommutativityGeneralDelta}), a NESS\ is singular
on a critical one-dimensional manifold $  \{\mbox{any}\ \Delta,\Gamma^{-1}=0,h=-1,g=-1\}$. According to
(\ref{NonCommutativityDelta1}),  further singularities appear for two special values of the anisotropy, inside
the critical manifold
 $\{\Delta=\pm1,\Gamma^{-1}=0,h=-1,g=-1\}$, engendering  a zero-dimensional
submanifold of the critical manifold. Thus, a hierarchy of singularities is formed.
It is quite remarkable that such hierarchical singularities
emerge without performing the thermodynamic limit $N\rightarrow
\infty$. In fact, as shown in Appendix \ref{app::Proof of hierarchical singularity in the NESS for N=4},
they can be explicitly detected already for  $N = 4$.
For $N=5$ we have found other singular manifolds, parametrized by $h,g,$ and $\Delta$.
For the sake of space, details of this case will be reported in a future publication.

The  appearance of the singularity at $h\rightarrow-1, g\rightarrow-1$ is
 a consequence of the  additional symmetry (\ref{SymmetryGlobal}) at this point.
By direct inspection of the analytic formulae obtained  for $N = 3,\, 4,\, 5$,
we can guess the form of the {\sl  limit  state}
$\lim_{\Gamma\rightarrow\infty}\lim_{h\rightarrow1}\lim_{\Delta\rightarrow1}$
as a fully factorized one, namely
\begin{eqnarray}
&&
\lim_{\Gamma\rightarrow\infty}\lim_{h\rightarrow-1}\lim_{\Delta\rightarrow
1}\rho_{NESS}(N,h,\Delta,\Gamma) = \\
&&
\rho_{L}\left(  \frac{1}{3}\sigma^{x}
-\frac{1}{3}\sigma^{y}+\frac{1}{2}I\right)^{\otimes_{N-2}}\rho_{R}.
\nonumber
\label{Delta=1limitingState}%
\end{eqnarray}
Conversely, for generic $\Delta$ and odd $N\geq5$, the limit  state $\lim_{\Gamma\rightarrow\infty}\lim_{h\rightarrow
-1} \rho_{NESS}(N,h,\Delta,\Gamma)$  does not take a factorized form.
 Notice also that from making use of
Eqs (\ref{SymmetryEven}) and (\ref{SymmetryOdd}), we readily obtain also the
NESS limit  state for $\Delta\rightarrow-1$:

\begin{eqnarray}
&& \lim_{\Gamma\rightarrow\infty}\lim_{h\rightarrow-1}\lim_{\Delta\rightarrow
-1}\rho_{NESS}(N,h,\Delta,\Gamma) = \\
&& \rho_{L}\prod\limits_{i=2}^{N-1}
\otimes\left(  \left(  -1\right)  ^{i}\frac{1}{3}\left(  (-1)^{N}\sigma
^{x}+\sigma^{y}\right)  +\frac{1}{2}I\right)  \rho_{R}. \nonumber
\label{Delta=-1limitingState}
\end{eqnarray}

\section{Conclusion.}
\label{sec::Conclusion}

In this paper we extensively analyzed the properties of the NESS of open Heisenberg
spin chains, subject to the action of LME at their boundaries and of perturbing
magnetic fields at the near-boundary sites. The setup we deal with operates in
the Zeno regime, i.e. in the strong coupling limit, $\Gamma \to +\infty$ (see
Eq. (\ref{LME}) ).  Most of our
analytic and numeric calculations have been performed for relatively small values of
the chain size $N$. On the other hand, as a consequence of the local nature
of the reservoirs and of the perturbing magnetic fields, we conjecture that many
of these results could be extended to large finite values of $N$: the delicate
question of how they might be modified in the thermodynamic limit is still
open.  At the present level of standard computational power, the strategy
of performing large scale calculations to get any inference on such a limit
is  impractical, because the number of equations to be solved grows exponentially
with $N$.

Despite all of these limitations, the main outcome of our study is quite
unexpected:
by tuning  the near--boundary magnetic fields we can manipulate the NESS,
making it pass from a dark pure state (for a suitable choice of the value of the anisotropy
parameter  $\Delta$), to a fully uncorrelated mixed state at infinite temperature.

We have also discussed how this general scenario emerges in the
anisotropic, partially anisotropic and isotropic cases. The influence of
different alignment conditions imposed by the Lindblad reservoirs has been
extensively explored, together with the symmetries of the NESS and their
importance for engendering hierarchical singularities due to the noncommutativity
of different  limits,  performed on the model parameters.

A physically relevant point in our discussion concerns the possibility of performing
such a  manipulation of the NESS also for large but finite values of $\Gamma$:
numerical investigations confirm this expectation, thus opening interesting
perspectives of experimental investigations.

\textbf{Acknowledgements} VP acknowledges the Dipartimento di Fisica e
Astronomia, Universit\`a di Firenze, for partial support through a FIRB initiative.
M.S. acknowledges support from the
Ministero dell' Istruzione, dell' Universit\'a e della Ricerca
(MIUR) through a {\it Programma di Ricerca Scientifica di
Rilevante Interesse Nazionale} (PRIN)-2010 initiative. A substantial part of the manuscript
was written during  a workshop in Galileo Galilei Institute in Florence. We thank David Mukamel for
useful discussions. RL acknowledges the support and the kind hospitality of MPIPKS in Dresden, where part of this
manuscript was written.
\appendix

\section{Inverse of the Lindblad dissipators and secular conditions.}
\label{app::Inverse of the Lindblad dissipators and secular conditions}

 $\mathcal{L}_{L}$ and $\mathcal{L}_{R}$ are linear super-operators acting on a matrix $\rho$ as defined by Eqs (\ref{L_Left}) and (\ref{L_Right}).
In our case, each super-operator act locally on a
single qubit only. The eigen-basis $\{\phi_{R}^{\alpha}\}_{\alpha=1}^{4}$ of
$\mathcal{L}_{R}\phi_{R}^{\alpha}=\lambda_{\alpha}\phi_{R}^{\alpha}$ is $\phi_{R}
=\{2\rho_{R},2\rho_{R} -I,
-sin\varphi_{R} \sigma^{x} + cos\varphi_{R} \sigma^{y},
cos\theta_R (cos\varphi_{R} \sigma^{x} + sin\varphi_{R} \sigma^{y})- sin \theta_R \sigma^{z}\},
$, with the respective eigenvalues  $\{\lambda_{\alpha
}\}=\{0,-1,-\frac{1}{2},-\frac{1}{2}\}$. Here $I$ is a 2$\times$2 unit matrix,
  $\sigma^{x},\sigma
^{y},\sigma^{z}$ are Pauli matrices, and $\rho_{R}$ is targeted spin opientation at the right
boundary.  Analogously, the eigen-basis and eigenvalues of the
eigenproblem $
\mathcal{L}_{L}\phi_{L}^{\beta}=\mu_{\beta}\phi_{L}^{\beta}$ are $\phi_{L}=\{2\rho_{L},2\rho_{L} -I,
-sin\varphi_{L} \sigma^{x} + cos\varphi_{L} \sigma^{y},
cos\theta_L (cos\varphi_{L} \sigma^{x} + sin\varphi_{L} \sigma^{y})- sin \theta_L \sigma^{z}\}$ and $\{\mu_{\beta}\}=\{0,-1,-\frac{1}{2},-\frac{1}{2}\}$,
where $\rho_{L} $ is the targeted spin opientation at the left
boundary. Since the bases
$\phi_{R}$ and $\phi_{L}$ are complete, any matrix $F$ acting in the
appropriate Hilbert space is expanded as
\begin{equation}
F=\sum\limits_{\alpha=1}^{4}\sum\limits_{\beta=1}^{4}\phi_{L}^{\beta}\otimes
F_{\beta\alpha}\otimes\phi_{R}^{\alpha}, \label{ro_expansion}%
\end{equation}
where $F_{\beta\alpha}$ are  $2^{N-2}\times 2^{N-2}$ matrices.
Indeed, let us introduce complementary bases
$\psi_L,\psi_R$ as
$\psi_{L,R}=\{I/2, \rho_{L,R} -I,
(-sin\varphi_{L,R} \sigma^{x} + cos\varphi_{L,R} \sigma^{y})/2,
(cos\theta_{L,R} (cos\varphi_{L,R} \sigma^{x} + sin\varphi_{L,R} \sigma^{y})- sin \theta_{L,R} \sigma^{z})/2\}$,
trace-orthonormal to the $\phi_{R},\phi_{L}$
respectively, $Tr(\psi_{R}^{\gamma}\phi_{R}^{\alpha})=\delta_{\alpha\gamma}$,
$Tr(\psi_{L}^{\gamma}\phi_{L}^{\beta})=\delta_{\beta\gamma}$. Then, the
coefficients of the expansion (\ref{ro_expansion}) are given by $F_{\beta
\alpha}=Tr_{1,N}((\psi_{L}^{\beta}\otimes I^{\otimes_{N-1}})F(I^{\otimes
_{N-1}}\otimes\psi_{R}^{\alpha}))$. On the other hand, in terms of the
expansion (\ref{ro_expansion}) the superoperator inverse $(
\mathcal{L}_{L}+\mathcal{L}_{R})^{-1}$ is simply
\begin{equation}
(\mathcal{L}_{L}+
\mathcal{L}_{R})^{-1}F=\sum\limits_{\alpha,\beta}\frac{1}{\lambda_{\alpha}+\mu_{\beta}%
}\phi_{L}^{\beta}\otimes F_{\beta\alpha}\otimes\phi_{R}^{\alpha}.
\label{LindbladInversion}
\end{equation}
The above sum contains a singular term with $\alpha=\beta
=1$, because $\lambda_{1}+\mu_{1}=0.$ To eliminate the singularity, one must
require $F_{11}=Tr_{1,N}F=0$, which generates the secular conditions
(\ref{SecularConditions}).

\section{Analytic treatment of $N=3$ case}
\label{app:Analytic treatment of $N=3$ case}

Here we prove the property (\ref{GibbsInfiniteTemperature}) for $N=3$, and demonstrate
a singularity of the NESS at a fixed value of local fields $h,g$.
Note that we treat case $N=3$ for simplicity and for demonstration purposes only;
Also for simplicity, we consider $XXZ$ Hamiltonian and perpendicular targeted polarizations in $XY$ plane
 $\vec{l_L}=(0,-1,0)$, $\vec{l_R}=(1,0,0)$,
\begin{equation}
H=H_{XXZ}-h\sigma_{2}^{y}+g\sigma_{N-1}^{x} \label{HamXXZ+LocalFields}%
\end{equation}
We have $\rho_{0}=\rho_{L}\otimes\left(  \frac{I}%
{2}+M_{0}\right)  \otimes\rho_{R}$ and
$\rho_{1}=2\mathcal{L}_{LR}^{-1}(i[H,\rho_{0}])+\rho_{L}\otimes M_{1}\otimes\rho_{R}$, with
$\rho_{L},\rho_{R}$ given by (\ref{RoL}),(\ref{RoR}), and $M_{0}=\sum
\alpha_{k}\sigma^{k}$, $M_{1}=\sum\beta_{k}\sigma^{k}$, where $\{\sigma
^{k}\}_{k=1}^{3}$ is a set of Pauli matrices, and $\alpha_{k},\beta_{k}$ are
unknowns. Secular conditions (\ref{SecularConditions}) at zero-th
order $k=0$ give a set of three equations%

\begin{align*}
(h+1)\alpha_{3}  &  =0\\
(g+1)\alpha_{3}  &  =0\\
(g+1)\alpha_{2}+(1+h)\alpha_{1}  &  =0,
\end{align*}
from which the $\rho_{0}$ cannot be completely determined. The secular
conditions (\ref{SecularConditions}) for $k=1$ provide missing relations,
\begin{align*}
-(h+1)\beta_{3}-2\left(  2\Delta^{2}+1\right)  \alpha_{1}+2\Delta &  =0\\
-(g+1)\beta_{3}-2\left(  2\Delta^{2}+1\right)  \alpha_{2}-2\Delta &  =0\\
(g+1)\beta_{2}+(h+1)\beta_{1}-4\alpha_{3}  &  =0
\end{align*}

from which $\rho_{0}$ can be readily found. Namely, if $h\neq-1$,$g\neq-1$,
then
\begin{align}
\alpha_{3}  &  =0\nonumber\\
\alpha_{1}  &  =(g+1)\frac{\Delta(g+h+2)}{\left(  2\Delta^{2}+1\right)
\left(  g^{2}+2g+h^{2}+2h+2\right)  }\label{N=3solution}\\
\alpha_{2}  &  =(-h-1)\frac{\Delta(g+h+2)}{\left(  2\Delta^{2}+1\right)
\left(  g^{2}+2g+h^{2}+2h+2\right)  }\nonumber
\end{align}

Observables of the system change nontrivially with $h, g$. In particular, the
current-like two-point correlation function $j^z_{12}=2\langle\sigma_{1}^{x}\sigma
_{2}^{y}-\sigma_{1}^{y}\sigma_{2}^{x}\rangle_{NESS}$ has the form

\begin{equation}
j^z_{12}=
4\alpha_{1}=4(g+1)\frac{\Delta(g+h+2)}{\left(2\Delta
^{2}+1\right)\left(g^{2}+2g+h^{2}+2h+2\right)}.
\end{equation}

Consequently, manipulating the $h,g$, one can change the sign of the above
correlation or make it vanish for all $\Delta$, for $2+g+h=0$. Moreover, for
$h=h_{cr}=-2-g$, all $\alpha_{k}=0$, see (\ref{N=3solution}), and we
recover (\ref{GibbsInfiniteTemperature}). If, however, $h=-1$,$g=-1$, then
the solution for $\alpha_{k}$ reads
\begin{align}
\alpha_{3}  &  =0\nonumber\\
\alpha_{1}  &  =-\alpha_{2}=\frac{\Delta}{2\Delta^{2}+1},
\label{NonCommutativityN3}%
\end{align}
manifesting a singularity of the NESS at the point $h=g=-1$ for any nonzero
$\Delta\neq0$, see also section \ref{sec::Noncommutativity}.

\section{Corrections to (\ref{GibbsInfiniteTemperature})  of the order $1/\Gamma$   }
\label{app::Proof of (ii)}

Here we show that the perturbation theory (\ref{Recurrence}) predicts
$M_{1}\neq0$ for arbitrary local fields $g,h$, if $M_{0}=0$. We restrict to $XXZ$ Hamiltonian
 $J_x=J_y=1$, and perpendicular boundary twisting in
the  $XY$--plane,  ${\vec l}_L =(0,-1,0)$, ${\vec l}_R =(1,0,0)$.

Let us set $\rho_{0}=\rho_{L}\left(  \frac{1}{2}I\right)  ^{\otimes_{N-2}}%
\rho_{R}$ as predicted by (\ref{GibbsInfiniteTemperature}) for critical values
of the local field. We then obtain, in the zeroth order of perturbation
\begin{eqnarray}
Q  &=& i[H,\rho_{0}] = i [h_{1,2} + h_{N-1,N},\rho_{0}]= \\
&=& \frac{1}{2^{N-2}}\left(  K_{XZ}\otimes I^{\otimes_{N-3}}\otimes\rho
_{R}-\rho_{L}\otimes I^{\otimes_{N-3}}\otimes K_{ZY}\right), \nonumber
\end{eqnarray}
where $K_{\alpha\beta}=-\Delta\sigma^{\alpha}\otimes\sigma^{\beta}%
+\sigma^{\beta}\otimes\sigma^{\alpha}$, and $h_{k,k+1}$ is the local
Hamiltonian term, $h_{k,k+1}=\sigma_{k}^{x}\sigma_{k+1}^{x}+\sigma_{k}%
^{y}\sigma_{k+1}^{y}+\Delta\sigma_{k}^{z}\sigma_{k+1}^{z}$. The secular
conditions $Tr_{1,N}Q=0$ are trivially satisfied. Noting that $Q$ has the
property $\frac{1}{2}\mathcal{L}_{LR}Q=-Q$, we obtain from (\ref{Recurrence0}) and (\ref{Recurrence}) the first
order correction to $\rho_{0}$
\[
\rho_{1}=-Q+\rho_{L}\otimes M_{1}\otimes\rho_{R}\text{.}%
\]
Let us assume that $M_{1}=0$. Then, in the second order of perturbation
theory, we have
\begin{eqnarray}
&& i [H,\rho_{1}] = -i [H,Q] = \\
&& -i[h_{12}+h_{23}+h\sigma_{2}^{y}+g\sigma_{N-1}%
^{x}+h_{N-2,N-1}+h_{N-1,N},Q].  \nonumber
\end{eqnarray}
After some calculations we obtain%
\begin{eqnarray}
&& i [H,\rho_{1}] = R+ const \times \\
&& \Delta(-I\otimes\sigma^{y}\otimes I^{\otimes_{N-3}%
}\otimes\rho_{R}+\rho_{L}\otimes I^{\otimes_{N-3}}\otimes\sigma^{x}\otimes
I),
\nonumber
\end{eqnarray}
where the unwanted secular terms are written out explicitly, and $Tr_{1,N}%
R=0$. The unwanted terms proportional to $\Delta$ do not depend on $h,g$. For
any $\Delta\neq0$ the secular conditions $Tr_{1,N}[H,\rho_{1}]=0$ cannot be
satisfied. This contradiction shows that $M_{1}\neq0$ for any  $\Delta\neq 0$.

\section{Hierachical singularity in the NESS for $N=4$}
\label{app::Proof of hierarchical singularity in the NESS for N=4}
Here we restrict to $XXZ$ Hamiltonian with $J_x=J_y=1$, and
perpendicular boundary twisting in
the  $XY$--plane ${\vec l}_L =(0,-1,0)$, ${\vec l}_R =(1,0,0)$.
For $N=4$ we
have $30$ equations to satisfy from the secular conditions
(\ref{SecularConditions}) for $k=0,1$, and the set of variables $\{\alpha
_{ki}\},\{\beta_{ki}\}$ to determine the matrices $M_{0}=\sum_{k,i=0}%
^{^{\prime}3}\alpha_{ki}\sigma^{k}\otimes\sigma^{i}$, $M_{1}=\sum^{\prime
}\beta_{ki}\sigma^{k}\otimes\sigma^{i}$. The "prime" in the sum denotes the
absence of the terms $\alpha_{00},\beta_{00}$ since the matrices $M_{k}$ are
traceless. The matrices $\{\sigma^{0},\sigma^{1},\sigma^{2},\sigma^{3}\}=$
$\{I,\sigma^{x},\sigma^{y},\sigma^{z}\}$ are unit matrix and Pauli matrices. We do
not list here all  $30$ equations but just their solutions for different
values of parameters, obtained using Matematica. For $g=-h-2$ we have, in agreement with
(\ref{GibbsInfiniteTemperature}), $M_{0}=0$, while, out of $15$ coefficients
$\{\beta_{ki}\}$, only six are determined, namely
\begin{align}
\beta_{13}  &  =\beta_{32}=1,\nonumber\\
\beta_{03}  &  =\beta_{31}=\frac{1}{1+h},\label{N4singular}\\
\beta_{23}  &  =\beta_{31}=0,\nonumber
\end{align}
while other $\beta_{ki}$ (and therefore, the $M_{1}$) have to be determined at
the next order $k=2$ of the perturbation theory. From (\ref{N4singular}) it is
clear that the case $1+h=0$ has to be considered separately. In fact, for
$h=g=-1$ we obtain a different solution:  $M_{0}=0$,
while the coefficients $\{\beta_{ki}\}$ are
\begin{align}
\beta_{13}  &  =\beta_{32}=\frac{\Delta^{2}}{-1+\Delta^{2}},\nonumber\\
\beta_{23}  &  =\beta_{31}=\frac{\Delta}{-1+\Delta^{2}}%
,\label{N4DeltaSingularity}\\
\beta_{01}  &  =\beta_{02}=\beta_{10}=\beta_{20}=0,\nonumber
\end{align}
thus at $h=g=-1$\ we have a singularity in the first order of perturbative
expansion, in $M_{1}$. On the other hand, (\ref{N4DeltaSingularity}) for
$\Delta=1$ there is a singularity in $M_{1}$: we have to treat this case
separately. For $\Delta=1$ we find $M_{0}=\left(  \frac{1}{3}\sigma^{x}%
-\frac{1}{3}\sigma^{y}+\frac{1}{2}I\right)^{\otimes_{2}}$, in agreement
with (\ref{Delta=1limitingState}), while the set of $\beta_{ki}$ is
\begin{align*}
\beta_{03}  &  =\beta_{30}=\frac{1}{2}\\
\beta_{32}  &  =\beta_{13}=1\\
\beta_{31}  &  =\beta_{23}=0.
\end{align*}
So at $\Delta=1,h=g=-1$ we have a singularity in the zeroth order of
the perturbative expansion, at the level of $M_{0}$. Summarizing, for $N=4$ we
have $M_{0}=0$ on the two-dimensional manifold of the phase space
characterized by $\{\Delta$ arbitrary, $g=-h-2\}$, except at the point
$\{\Delta=1,h=g=-1\}$, where $M_{0}=\left(  \frac{1}{3}\sigma^{x}-\frac{1}%
{3}\sigma^{y}+\frac{1}{2}I\right)  ^{\otimes_{2}}$. On a one-dimensional
submanifold $\{\Delta\neq1$ , $g=h=-1\}$ there is a singularity in $M_{1}$.

\end{document}